      [1999/12/01 v1.4c Il Nuovo Cimento]
\documentclass{cimento}
\usepackage{graphicx}

\title{GRBs with optical afterglow and known redshift: a statistical study}

\author{
G.~Greco\from{ins:bologna},
D.~Bad'in\from{ins:sai},
G.~Beskin\from{ins:sao},
C.~Bartolini\from{ins:bologna},
S.~Karpov\from{ins:sao},
A.~Guarnieri\from{ins:bologna},
A.~Piccioni\from{ins:bologna},
\atque
A.~Biryukov\from{ins:sai}.
}
\shortauthor{Greco, G. et al}

\instlist{
  \inst{ins:bologna} Astronomy Department of Bologna University, Bologna, Italy
  \inst{ins:sai} Sternberg Astronomical Institute of Moscow State University,
  Moscow, Russia
  \inst{ins:sao} Special Astrophysical Observatory of Russian Academy of Sciences, Russia
}
\PACSes{
  \PACSit{95.55.Cs}{Ground-based ultraviolet, optical and infrared telescopes}
  \PACSit{98.70.Rz}{Gamma-ray sources; Gamma-ray bursts}
}
\begin{document}

\maketitle

\begin{abstract}

  We present a correlation between two intrinsic parameters of GRB optical afterglows.
  These are the isotropic luminosity at the maximum of the light curve
  ($L_{peak}$) and the time-integrated isotropic energy  ($E_{iso}$) radiated after the observed maximum.
  We test  the correlation between the logarithms of ($E_{iso}$) and ($L_{peak}$)
  and finally we value the effect of the different samples of GRBs in according with the first optical observation 
  reduced to proper time. 

\end{abstract}

We analized the $ R $-band afterglow light curves of  63 long-duration GRBs/XRFs 
detected during the time period from February 1997 to October 2006. 
We obtained a sample with 42 GRBs/XRFs in which the light curves are generally well-sampled 
until detection of  optical afterglow in the host galaxy system, 
the redshifts of all bursts are measured and the estimates of the 
optical extinction  in the source frame are available.
The photometric data of optical afterglow light curves, as well as the values of the redshift, 
the host magnitude, the spectral slope ($\beta$) and the intrinsic extinction ($A_{v}$), were compiled by publications and GCN.
Before performing a statistical analysis, the observational data were corrected for extinction
(both Galactic and intrinsic) and the flux contribution from the host galaxy was subtracted.
 Then we converted  the magnitudes to fluxes using  the normalization given by \cite{fukugita}.
To avoid an influence of any model assumptions about the light curve shapes
(i.e. re-brightening episodes, jet-breaks) we decided to integrate them numerically by means of trapezoids,
without the need to extrapolate the flux to some epoch.
Finally the GRB sample was divided into four distinct sub-samples in according with the first optical 
detection ($t_{first}$) in the rest frame of the source. The series of temporal ranges $ (i-iv) $, the Pearson correlation 
coefficients ($ R $), the slopes ($\alpha$), and the number
of the GRBs/XRFs  in the different sub-samples  ($ N $) are given in Table I. 
  
\begin{figure}
\includegraphics[width=6.5cm]{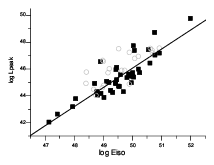}
\includegraphics[width=6.5cm]{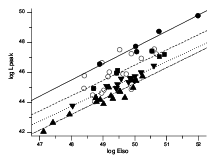}
\caption{ Left panel -- The best fit for the completed sample.
 Right panel -- The best fit for the four sub-samples in according with the
 first optical observation reduced to proper time: 
(i) solid line,  (ii) dash line,  (iii) dot line, (iv) dash-dot line.
 Open circles correspond to the bursts for
which there is no information on the amount of extinction in the host.
.
} \label{fig_1}
\end{figure}

\begin{table}
\caption{Correlation parameters for various samples of GRBs/XRFs}
\label{params}
\label{table_1}
\begin{narrowtabular}{1cm}{ll|c|c|c}
\hline
\multicolumn{2}{c|}{Sample features} & $\alpha$ & $ R $ & $ N $  \\
\hline

&  completed sample     & 1.43 $\pm$ 0.13 & 0.87 & 42\\
(i) 	 & $t_{first}$ $< 75 sec$	& 1.08 $\pm$ 0.11 & 0.97 & 5\\
(ii) 	 & $ 90 sec <$ $t_{first}$  $< 350 sec$ & 1.02 $\pm$ 0.14 & 0.96 & 6\\
(iii) 	 & $500 sec<$ $t_{first}$ $< 18.000 sec$	& 0.99 $\pm$ 0.08 & 0.96 & 13\\
(iv) 	 & $ 19.000 sec<$ $t_{first}$ $< 100.000 sec$	& 1.04 $\pm$ 0.08 & 0.95 & 18\\

\hline
\end{narrowtabular}
\end{table}

The previous approach reduces the scatter of the data points around the correlation.
For the bursts whose ($\beta$) and ($A_{v}$) are not available yet (marked by open circles in fig. 1)
we assumed $\beta$ = 1 and $A_{v}$ = 0. We did not utilize  them for the estimate of the correlation parameters  
given in table I. 
Detected correlation for 42 GRBs/XRFs confirms our previous results for 22 ones \cite{bartolini}.
We believe  using this phenomenological relationship 
 will enable us  to constrain the values of the
absorption in host galaxy system and the energetic ratios 
among the different physical processes  involved in the observed optical light curves.
Because of the small size of this sample, however, this conclusion is tentative.
Future observations  will provide a larger and less biased sample 
to test the  $L_{peak}$-$E_{iso}$ correlation in optical range.

\acknowledgments

This work was supported by grants of RFBR (No. 04-02-17555), Bologna University
(Progetti Pluriennali 2003)

\end{document}